\documentclass[conference]{IEEEtran}
\IEEEoverridecommandlockouts
\usepackage{cite}
\usepackage{amsmath,amssymb,amsfonts}
\usepackage{algorithmic}
\usepackage{graphicx}
\usepackage{textcomp}
\usepackage{xurl}
\usepackage{xcolor}
\usepackage{hyperref}
\usepackage{comment}
\usepackage{makecell}
\usepackage{booktabs}
\usepackage{multirow}
\usepackage{tabularx}
\usepackage{tcolorbox} 
\usepackage{array}
\usepackage{orcidlink}
\usepackage{subcaption}
\def\BibTeX{{\rm B\kern-.05em{\sc i\kern-.025em b}\kern-.08em
    T\kern-.1667em\lower.7ex\hbox{E}\kern-.125emX}}
\begin{document}

\title{Optimizing ROS~2 Communication for \\ Wireless Robotic Systems}

\author{
  \IEEEauthorblockN{%
    Sanghoon~Lee\,\orcidlink{0000-0002-8160-8952},
    Taehun~Kim\,\orcidlink{0009-0005-5193-553X},
    Jiyeong~Chae\,\orcidlink{0000-0003-3014-7559},
    and Kyung-Joon~Park\,\orcidlink{0000-0003-4807-6461}}
  \IEEEauthorblockA{Department of Electrical Engineering and Computer Science, DGIST, Daegu, Republic of Korea\\
  Email: \{leesh2913, taehun, cowldud3, kjp\}@dgist.ac.kr}
}

\maketitle

\begin{abstract}
Wireless transmission of large payloads, such as high-resolution images and LiDAR point clouds, is a major bottleneck in ROS~2, the leading open-source robotics middleware. 
The default Data Distribution Service (DDS) communication stack in ROS~2 exhibits significant performance degradation over lossy wireless links.
Despite the widespread use of ROS~2, the underlying causes of these wireless communication challenges remain unexplored. 
In this paper, we present the first in-depth network-layer analysis of ROS~2's DDS stack under wireless conditions with large payloads. We identify the following three key issues: excessive IP fragmentation, inefficient retransmission timing, and congestive buffer bursts.
To address these issues, we propose a lightweight and fully compatible DDS optimization framework that tunes communication parameters based on link and payload characteristics. 
Our solution can be seamlessly applied through the standard ROS~2 application interface via simple XML-based QoS configuration, requiring no protocol modifications, no additional components, and virtually no integration efforts. 
Extensive experiments across various wireless scenarios demonstrate that our framework successfully delivers large payloads in conditions where existing DDS modes fail, while maintaining low end-to-end latency.
\end{abstract}


\section{Introduction}
\label{sec1}
Modern robotic systems rely on high-resolution sensors--such as LiDARs, RGB cameras, and depth cameras--and often integrate edge/cloud offloading to enable intelligent functionality. As a result, the reliable wireless transmission of large-payload data has become essential.
In real-world environments where multiple robots operate simultaneously, concurrent data exchange is prone to latency, packet loss, and jitter. These communication issues can significantly impair decision-making and control in robots. 
Therefore, ensuring reliable and low-latency wireless communication is critical for the success of key applications such as autonomous navigation, teleoperation, and multi-robot collaboration.

To meet these communication requirements, the Robot Operating System 2~(ROS~2)~\cite{macenski2022robot} has emerged as the de facto standard middleware framework in robotics. ROS~2 uses the Data Distribution Service (DDS) as its core communication layer~\cite{8607261}, which employs a brokerless, peer-to-peer architecture based on the Real-Time Publish-Subscribe (RTPS) protocol. DDS provides fine-grained Quality of Service (QoS) configurations to support various robotic workloads, including sensor data streaming, control commands, and state feedback~\cite{kang2012rdds}.

\begin{comment}
The DDS-based communication architecture in ROS~2 faces significant challenges in reliably transmitting large data payloads over wireless networks. %[Bellavista; Kato]  %% what kind of limitation??????
This problem is well recognized within the ROS community. For example, Fast DDS--the default DDS implementation in ROS~2--offers an optional Transmission Control Protocol (TCP) transport mode to mitigate this problem. However, this heuristic approach is specific to Fast DDS, lacks interoperability with standard ROS nodes, and deviates from the core principles of DDS as a real-time distributed system. To the best of our knowledge, no general-purpose solution currently exists in ROS~2 for reliably and efficiently transmitting large payloads over wireless links.
To facilitate the adoption of edge and cloud-based applications in robotic networks, it is essential to enable reliable transmission of large-payload data even over wireless links. 
\end{comment}

The DDS-based communication architecture in ROS~2 struggles to reliably transmit large payloads over wireless networks, which is a well-known issue in the ROS community.
Fast DDS, the default DDS implementation in ROS~2, offers an optional Transmission Control Protocol (TCP)-based transport mode as a workaround. 
However, this solution is specific to Fast DDS, lacks interoperability with standard ROS nodes, and deviates from the core principles of DDS as a real-time distributed system. 
To date, no general-purpose mechanism exists in ROS~2 for robust and efficient transmission of large payloads over unreliable wireless links. Addressing this gap is essential for supporting the growing demand for edge and cloud-based robotic applications, which typically rely on the wireless transmission of large-payload data. 

In this paper, we analyze the DDS-based ROS~2 communication stack in the context of wireless transmission failures, with a particular focus on its limitations in reliable large-payload transfer.
Based on this analysis, we propose a lightweight DDS optimization framework that significantly improves wireless transmission performance and establishes the foundation for reliable edge-driven robot applications.
The main contributions of this paper are summarized as follows:
\begin{itemize}
\item We analyze three key factors that lead to large-payload transmission failures in ROS~2 over wireless networks: IP fragmentation, retransmission timing, and buffer burst.
We quantitatively evaluate each factor's impact on transmission reliability. 

\item We present a configurable DDS optimization framework designed to significantly enhance wireless transmission performance for large payloads
without requiring low-level modifications. %, which ensures ease of integration and deployment. 
%Our method requires no low-level system modifications, ensuring ease of integration and deployment.
Fully compatible with existing ROS~2 systems, the proposed framework can be seamlessly applied through the standard ROS~2 application interface using simple XML-based QoS configuration files. 
%This design allows developers to optimize communication performance without altering application code or core middleware components.

\item We conduct a comprehensive experimental study across various wireless channel conditions and payloads derived from standard ROS~2 message types. 
The results show that the proposed framework reliably delivers large payloads even in scenarios where existing DDS modes fail, while maintaining low end-to-end latency.
\end{itemize}

The remainder of this paper is organized as follows. 
Section~\ref{sec2} reviews related work. 
Section~\ref{sec3} describes ROS~2 DDS communication mechanism, covering network layers from the application level to the physical link. 
Section~\ref{sec4} analyzes the causes of large-payload transmission failures over wireless networks and introduces our DDS optimization method. 
Section~\ref{sec5} evaluates its performance in real-world wireless environments. Finally, Section~\ref{sec6} concludes the paper.

\section{Related Work}
\label{sec2}
ROS~2 is an open-source platform that has become the de facto standard for robotic software development, enabling seamless integration of various sensors, actuators, and control modules into robotic applications.
ROS~2 applications are developed using the high-level API called the ROS Client Library (RCL), which connects to the DDS communication layer through the ROS Middleware (RMW). 
In ROS~2, all message exchange is handled by DDS, which ensures both real-time performance and reliable data transfer.

DDS serves as the middleware that bridges ROS applications with the underlying operating system. 
In ROS applications, the fundamental unit of communication is the node.
DDS provides a distributed publish/subscribe model in which nodes exchange data directly without a central broker. 
Each node discovers and connects to peers automatically through a built-in discovery protocol. 
DDS also offers more than 15 QoS policies, including Reliability, Deadline, and Durability, allowing adjustment of performance parameters such as latency and reliability.
These features make DDS particularly well-suited for robotic applications that require real-time control and high-frequency sensor communication~\cite{wytrkebowicz2021messaging}.

DDS adopts the User Datagram Protocol (UDP) as its default transport layer, primarily for two reasons. 
First, UDP supports multicast transmission, which allows data to be delivered to multiple nodes simultaneously. 
This significantly improves throughput compared to unicast-based delivery and reduces both network and computational overhead~\cite{kang2020evaluating}. 
Second, UDP offers lower latency and minimal protocol overhead. 
In contrast, TCP introduces delays due to connection setup, congestion control, and acknowledgment mechanisms, which can hinder real-time performance in robotic systems. 
UDP's simpler transmission model avoids these complexities, making it more suitable for latency-sensitive communication.
Prior studies have empirically demonstrated that UDP-based middleware consistently outperforms TCP in terms of latency, message size handling, and protocol overhead, in both wired and wireless environments~\cite{liang2023performance, moraes2019performance, thangavel2014performance}. 
For these reasons, DDS relies on UDP as the default transport protocol. 
Instead, DDS implements essential functions such as session management, flow control, and retransmission at the middleware level.

Despite these advantages, current DDS implementations show clear limitations in large-payload transmissions. 
Experimental studies by~\cite{bellavista2013data} showed that, in a 100~Mb/s wired LAN, two DDS implementations exhibited sharp increases in latency and overhead as payload size increased.
These performance issues were primarily caused by fragmentation and per-fragment acknowledgment processing~\cite{bellavista2013data}.
Similarly, Kato \emph{et al.}\cite{maruyama2016exploring} evaluated multiple DDS implementations and observed sharp latency spikes and transmission failures when payloads exceeded 64~KB.
Notably, they reported process crashes when sending messages of 128~KB or more\cite{maruyama2016exploring}.
These findings provide empirical evidence that fragmentation and acknowledgment overhead already constitute performance bottlenecks, even in wired environments.

This problem becomes more severe over wireless links, where packet loss and channel variability are inherent. 
Peeck \emph{et al.}~\cite{peeck2021middleware} evaluated the reliability of standard DDS by transmitting 20~KB image samples over an IEEE 802.11p network.
They observed that the protocol would stall upon the loss of a single fragment, remaining inactive until the next scheduled retransmission trigger. 
As a result, the system failed to meet the sample deadline of 100~ms, even under very low frame error rate (FER). 
When the FER increased to 50\% (corresponding to a packet error rate (PER) of approximately 10\%), the deadline miss rate rose sharply. 
These results indicate that standard DDS is fundamentally incapable of supporting real-time, large-payload transmission over wireless networks~\cite{peeck2021middleware}.

The ROS community has also acknowledged this limitation. eProsima Fast DDS, the default DDS implementation in ROS~2, provides a mode called LARGE\_DATA to mitigate the issues that arise when transmitting large payloads over lossy wireless links~\cite{EduPonz2024_LargeData}.
However, this mode transmits messages over TCP (or shared memory) instead of UDP, making it a vendor-specific extension rather than part of the standard DDS specification. 
Consequently, it lacks interoperability with standard UDP-based nodes and sacrifices the low-latency advantages associated with UDP.

ROS~2 users typically expect the system to `just work' without requiring low-level modifications.
However, the current DDS-based communication in ROS~2 frequently fails to transfer large payloads over wireless networks. 
This study investigates such failures and identifies three primary causes underlying large-payload transmission in ROS~2 communication.
Based on this analysis, we propose a DDS optimization method that enhances reliability through adjustments to standard QoS parameters that are fully accessible via the user-level configuration interface, without introducing any additional mechanisms or modifying the transport protocol.

\section{Preliminaries}
\label{sec3}
This section describes the end-to-end flow of ROS~2 communication, including the procedures in the lower layers, and explains the reliability mechanisms adopted in DDS.
\subsection{ROS~2 Communication Stack}
\label{sec3-1}
In ROS~2, message transfer is abstracted as a publish/subscribe interaction over a topic, which is a logical channel identified by a name and message type, and compatible QoS policies. 
Users interact with the system at the application level using the publish() API to send message samples and the subscribe() API to receive them, without needing to be aware of the lower-layer details. 
The entity that sends messages is called a \emph{Publisher}, while the entity that receives them is referred to as a \emph{Subscriber}. 
A \emph{Node} in ROS~2 represents an application-level execution unit, and multiple Publishers and Subscribers can be instantiated within a single node. 

\begin{figure}[t]
    \centering
    \includegraphics[width=\linewidth]{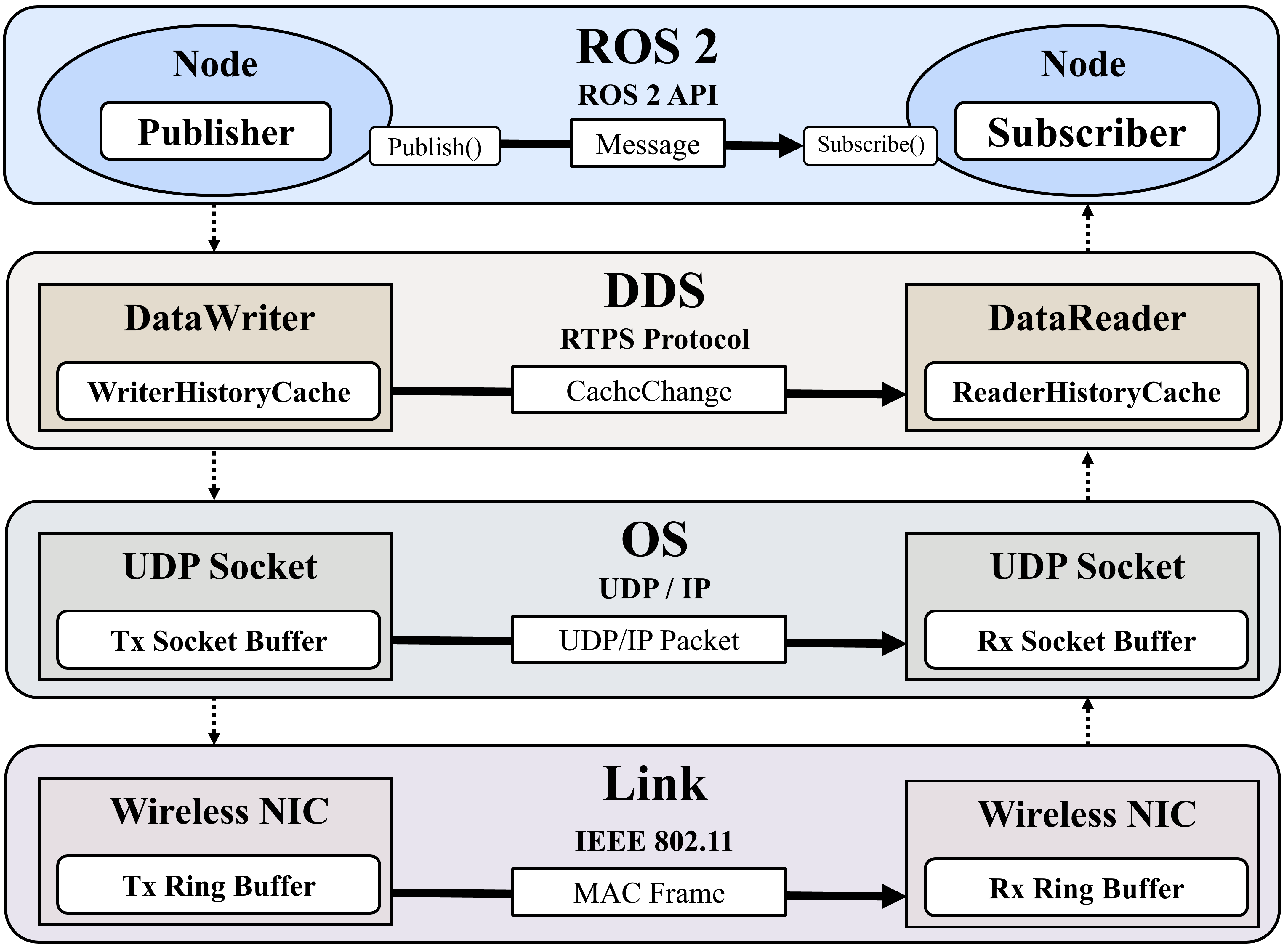}
    \caption{ROS~2 end-to-end communication stack.}
    \label{fig:ROS2 Stack}
\end{figure}

Data exchange between two nodes sharing a single topic follows the layered architecture illustrated in Fig.~\ref{fig:ROS2 Stack}.
ROS~2 DDS communication proceeds through a hierarchical process consisting of (1) the ROS~2 application layer, (2) the DDS middleware layer, (3) the OS network layer, and (4) the physical link layer.
At the top layer, user-invoked high-level APIs such as publish() and subscribe() are implemented via language-specific client libraries (e.g., rclcpp, rclpy) and passed to the RCL.
The RCL interacts with the selected DDS implementation (e.g., Fast DDS, Cyclone DDS) through the RMW interface.
Within the RMW implementation, each ROS communication entity is mapped one-to-one to a corresponding DDS entity. 
Specifically, a ROS Publisher is mapped to a DDS DataWriter, and a ROS Subscription is mapped to a DDS DataReader.

DDS serves as middleware that bridges ROS applications to the underlying OS network layer, transporting serialized messages over the RTPS protocol while providing discovery (session establishment) and reliable transmission.
Each DataWriter and DataReader maintains a dedicated HistoryCache, specifically, a WriterHistoryCache and a ReaderHistoryCache.
When publish() is invoked in a ROS application, the message sample is first serialized into a CacheChange payload using the CDR or XCDR2 format.
This payload is then stored in the WriterHistoryCache and transmitted via the RTPS protocol.
Conversely, received samples are placed into the ReaderHistoryCache via RTPS and delivered to the application through the subscribe() callback.

In essence, data exchange between a Publisher and Subscriber is realized by the RTPS protocol synchronizing changes from the WriterHistoryCache to the ReaderHistoryCache as faithfully as possible.
Thus, the HistoryCache acts as both a buffer and an interface between DDS entities and the RTPS transport layer.

Through the RTPS protocol, the serialized payload is encapsulated with an RTPS message header and split into one or more RTPS messages, depending on its size. 
These messages are then passed down to the transport layer.
Depending on the implementation, the transport layer may support UDP, TCP, or shared memory (SHM). However, the OMG RTPS standard assumes UDP/IP as the default transport, and most DDS implementations follow this convention.
As a result, under default settings, RTPS messages are transmitted via UDP sockets to the OS network layer.

RTPS messages are encapsulated within UDP datagrams and passed through the OS network stack. 
These datagrams are first queued in the UDP socket buffer and then forwarded to the IP layer. 
If a packet exceeds the network's Maximum Transmission Unit (MTU), the IP layer fragments it into multiple IP fragments. 
Each fragment is subsequently placed in the Network Interface Card (NIC)'s transmission ring buffer, where it is encapsulated in a link-layer frame and transmitted over the wireless interface (e.g., Wi-Fi). 
At the receiver, the IP layer reassembles the fragments to reconstruct the original UDP datagram. 
This datagram is parsed into an RTPS message, which is then stored in the ReaderHistoryCache as a CacheChange. 
Once the HistoryCache is updated, the new message sample is delivered to the ROS~2 application.

\subsection{Reliability Mechanism of DDS}
\label{sec3-2}
\begin{figure}
    \centering
    \includegraphics[width=\linewidth]{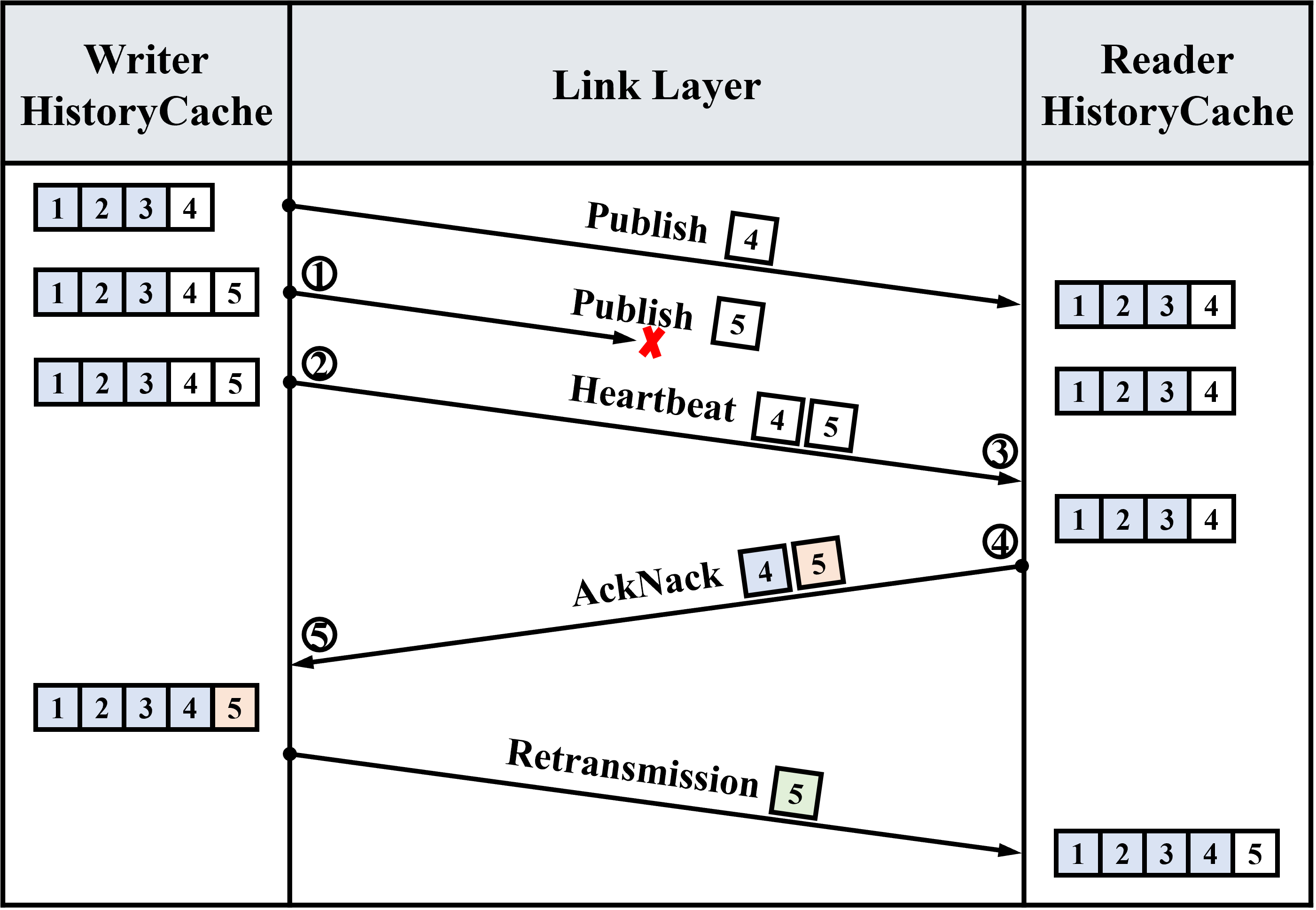}
    \caption{Retransmission mechanism of DDS}
    \label{fig:retran_mech}
\end{figure}

When transmitting over wireless links, packet loss occurs due to wireless channel impairments such as radio interference and channel variability.
Because UDP provides no built-in reliability, DDS ensures dependable communication using a retransmission mechanism based on Heartbeat and AckNack messages, configured via QoS policies.
The operational flow is illustrated in Fig.~\ref{fig:retran_mech}.
\begin{enumerate}
    \item {\textbf{[Publish]} When a Publisher publishes a message sample, DataWriter stores it in the WriterHistoryCache.} %in the form of a CacheChange.}

    \item {\textbf{[Heartbeat Transmission]} The DataWriter periodically sends Heartbeat RTPS control messages to all DataReaders subscribed to the same topic. The heartbeat interval can be configured via QoS parameters. Each Heartbeat includes the minimum and maximum sequence numbers currently stored in the WriterHistoryCache.}

    \item {\textbf{[Heartbeat Reception]} When the DataReader receives a Heartbeat, it compares its ReaderHistoryCache against the sequence range to identify any missing samples.}

    \item {\textbf{[AckNack Transmission]} The DataReader sends an AckNack message to the DataWriter. If samples are missing, their sequence numbers are encoded in a bitmap. Otherwise, a positive Ack is indicated.}

    \item {\textbf{[AckNack Reception]} When the DataWriter receives an AckNack message from a DataReader, it examines the reported sequence numbers to identify samples that require retransmission or acknowledgment.}

    \item {\textbf{[Retransmission]} The DataWriter then selectively retransmits the missing RTPS messages. Even after retransmission, CacheChange entries remain in the WriterHistoryCache until all subscribed DataReaders acknowledge them.}
\end{enumerate}

\section{DDS Optimization for Wireless Large-Payload Transfers}
\label{sec4}
This section proposes a DDS optimization framework for improving large-payload transmission performance over wireless networks. 
The transmission problem in ROS~2 is first defined, and three key failure causes are identified: (1) IP fragmentation, (2) retransmission timing, and (3) buffer burst. 
Then, we present a practical optimization approach that requires no additional implementation or changes to the transport mode.
The proposed method can be easily applied using XML profiles supported by existing DDS vendors.

\subsection{Problem Description}
\label{sec4-1}
In ROS~2, the problem of large-payload data transmission over wireless networks can be described as follows:
Consider two hosts with different MAC and IP addresses, each running either a Publisher or a Subscriber on the same topic.
This topic is assumed to carry data types such as integer arrays, images, and point clouds, and it is configured with a strictly reliable QoS policy.
Here, strictly reliable communication implies that every message sample must be delivered without loss.
To achieve this, the reliability QoS must be set to RELIABLE, and the history QoS must be set to KEEP\_ALL.

Suppose the Publisher transmits message samples of size $u$~B at a rate of $r$~Hz on the given topic. 
Then, the application-layer transmission rate $R_{\text{pub}}$ can be expressed as
\begin{equation} R_{\text{pub}} = r \cdot u \qquad [\mathrm{B/s}]. \end{equation}

As described in Section~\ref{sec3-1}, the data flow in ROS~2 proceeds in the following order: Application layer $\rightarrow$ DDS layer $\rightarrow$ OS layer $\rightarrow$ Link layer.
On the receiving side, data is processed in the reverse direction.
At each layer or interface boundary, bottlenecks may occur due to bandwidth limits, latency, or buffer constraints.
Thus, even if the upper layers operate fast enough, the overall transmission throughput can still be limited if any lower layer becomes saturated.
Accordingly, the maximum achievable throughput $T_{\max}$ is determined by the minimum capacity among all layer boundaries:
\begin{equation}
T_{\max}
=\min\!\bigl(
T_{\text{App}\rightarrow\text{DDS}},
T_{\text{DDS}\rightarrow\text{OS}},
T_{\text{OS}\rightarrow\text{Link}}
\bigr),
\end{equation}
where $T_{\text{App}\rightarrow\text{DDS}}$ denotes the throughput at which message samples invoked by the application API can be written into the HistoryCache.
$T_{\text{DDS}\rightarrow\text{OS}}$ refers to the throughput at which CacheChange entries are flushed from the HistoryCache to the OS-level UDP socket by the RTPS protocol.
$T_{\text{OS}\rightarrow\text{Link}}$ represents the effective bandwidth from the OS kernel space to the physical link, as packets are transmitted via NIC. % (TX ring). This rate is influenced by both the kernel-to-NIC handoff speed and the physical link capacity.

If the transmission rate $R$ persistently exceeds any of these throughput limits $T$, it leads to back-pressure toward the upper layers or buffer overflows, causing sample or packet drops and resulting in increased end-to-end latency. 
In most ROS~2 communication, it has been empirically observed that both $T_{\text{App}\rightarrow\text{DDS}}$ and $T_{\text{DDS}\rightarrow\text{OS}}$ significantly exceed $T_{\text{OS}\rightarrow\text{Link}}$~\cite{maruyama2016exploring}.
Benchmark results show that Fast DDS running on ROS~2 can internally publish messages with 0.5~MB per sample at nearly 550--660~Hz, corresponding to an internal throughput of about 275--330~MB/s in the UDP mode~\cite{eprosima_fastrtps_perf}.

In contrast, the practical throughput of typical network links is far lower:
100~Mb/s Ethernet typically delivers around 12.5~MB/s~\cite{6778661},
2.4~GHz Wi-Fi links sustain only 20--30~Mb/s (2.5--3.75~MB/s), 
while 5~GHz Wi-Fi offers approximately 60--80~Mb/s (7.5--10~MB/s)~\cite{8215954}.
These clearly illustrate that $T_{\max}$ = $T_{\text{OS}\rightarrow\text{Link}}$. 
Accordingly, we focus on minimizing average latency and jitter for large-payload data transmission under constrained $T_{\text{OS}\rightarrow\text{Link}}$ conditions.

\subsection{Characteristics of Wireless Links}
\label{sec4-2}
The fundamental differences between wireless and wired links lie in their packet loss characteristics due to channel variability. 
Wireless links, particularly Wi-Fi, inherently suffer from probabilistic packet loss due to the nature of Carrier Sense Multiple Access with Collision Avoidance (CSMA/CA). Furthermore, the available bandwidth frequently fluctuates due to interference, congestion, and varying channel fading or shadowing.

\begin{table}[h]
\centering
\caption{Message delivery rate (\%) by payload size}
\label{tab:table1}
\begin{tabularx}{\linewidth}{>{\centering\arraybackslash}m{2.3cm} *{10}{>{\centering\arraybackslash}X}}
\toprule
\textbf{\makecell{Payload (KB)\\Link Type}} & \textbf{33} & \textbf{66} & \textbf{99} & \textbf{132} & \textbf{165} & \textbf{198} & \textbf{231} & \textbf{264} & \textbf{297} & \textbf{330} \\
\midrule
Ethernet               & 100 & 100 & 100 & 100 & 100 & 100 & 100 & 100 & 100 & 100 \\
Wi-Fi                  & 99  & 98  & 98  & 97  & 98  & 97  & 97  & 97  & 95  & 97  \\
Wi-Fi (PER 0.1\%)      & 97  & 94  & 90  & 88  & 85  & 84  & 82  & 81  & 78  & 73  \\
Wi-Fi (PER 1\%)        & 77  & 61  & 47  & 37  & 28  & 23  & 18  & 13  & 10  & 9   \\
\bottomrule
\end{tabularx}
\end{table}
Table~\ref{tab:table1} presents the message delivery rate under unreliable transmission across various physical link conditions in ROS~2 Humble. 
In this experiment, 10,000 messages were transmitted for each payload size ranging from 33~KB to 330~KB under different link types. 
Ethernet consistently achieved a delivery rate of 100\%  regardless of the payload size, whereas Wi-Fi exhibited a message loss rate of approximately 1--3\%. 
Message delivery rate severely degraded as the payload size increases, particularly under lossy link conditions.
These results highlight a fundamental limitation of wireless links in large-payload transmissions.

\subsection{IP Fragmentation}
\label{sec4-3}
The reduced message delivery rate with larger payloads is due to packet-level fragmentation inherent in DDS over wireless links.
In DDS, a large message sample of size $u$ is first divided into $N_{RT} = \lceil u / M \rceil$ RTPS messages, where $M$ is the maximum RTPS payload (typically 64~KB). 
Each RTPS message is then further fragmented into $N_{\mathrm{IP}} = \lceil M / 1500 \rceil$ IP packets. 
Due to this hierarchical fragmentation, the successful delivery of a message sample requires the lossless transmission of all the corresponding IP packets. 
Assuming the delivery rate of a single IP packet is $p$, the probability that the entire RTPS message is received without error is $q = p^{N_{\mathrm{IP}}}$. 
Consequently, the delivery rate of the full sample decreases rapidly as $u$ increases, even for moderate packet error rates. 
% This effect explains the sharp drop in delivery performance for large payloads observed in Table~\ref{tab:table1}.

When a sample is lost, DDS ensures reliability through the retransmission mechanism detailed in Section~\ref{sec3-2}.
This mechanism is periodically activated by Heartbeat and AckNack exchanges, which occur at a fixed frequency of $n$~Hz, independently of the publish rate $r$.
Under this mechanism, the number of samples accumulated between two retransmission rounds follows a discrete-time staircase pattern.
Specifically, the number of newly published samples available at the $k$-th retransmission round is given by
\[\Delta S_k = \left\lfloor \frac{r}{n} \cdot k \right\rfloor - \left\lfloor \frac{r}{n} \cdot (k - 1) \right\rfloor,\]
which results in a periodic burst of publish traffic.

Retransmission is only possible when both Heartbeat and AckNack messages are successfully exchanged, which occurs with probability $p^2$. 
Accordingly, the total traffic per retransmission round consists of two components: (i) the newly published data $\Delta S_k \cdot u$, and (ii) the retransmission of data that failed in the previous retransmission round.
Therefore, the total transmission in the $k$-th round is given by
\[ X_k = \Delta S_k \cdot u + p^2 (1 - q) \cdot X_{k-1}. \]

Because each round spans $1/n$ seconds and the publish rate is $r$, $\Delta S = \mathbb{E}[\Delta S_k] = r/n.$
Therefore, the total transmission per retransmission round converges to a steady-state value $X$ as $k \to \infty$ if $p^2(1-q)<1$.
\begin{equation}
\label{eq_traffic}
   X = \frac{r \cdot u}{n \left(1 - p^2 + p^2q \right)} = \frac{ R_{\text{pub}}}{n \left(1 - p^2 \left(1 - p^{N_{\mathrm{IP}}} \right) \right)}. 
\end{equation}

\begin{figure}[t]
    \centering
    \includegraphics[width=\linewidth]{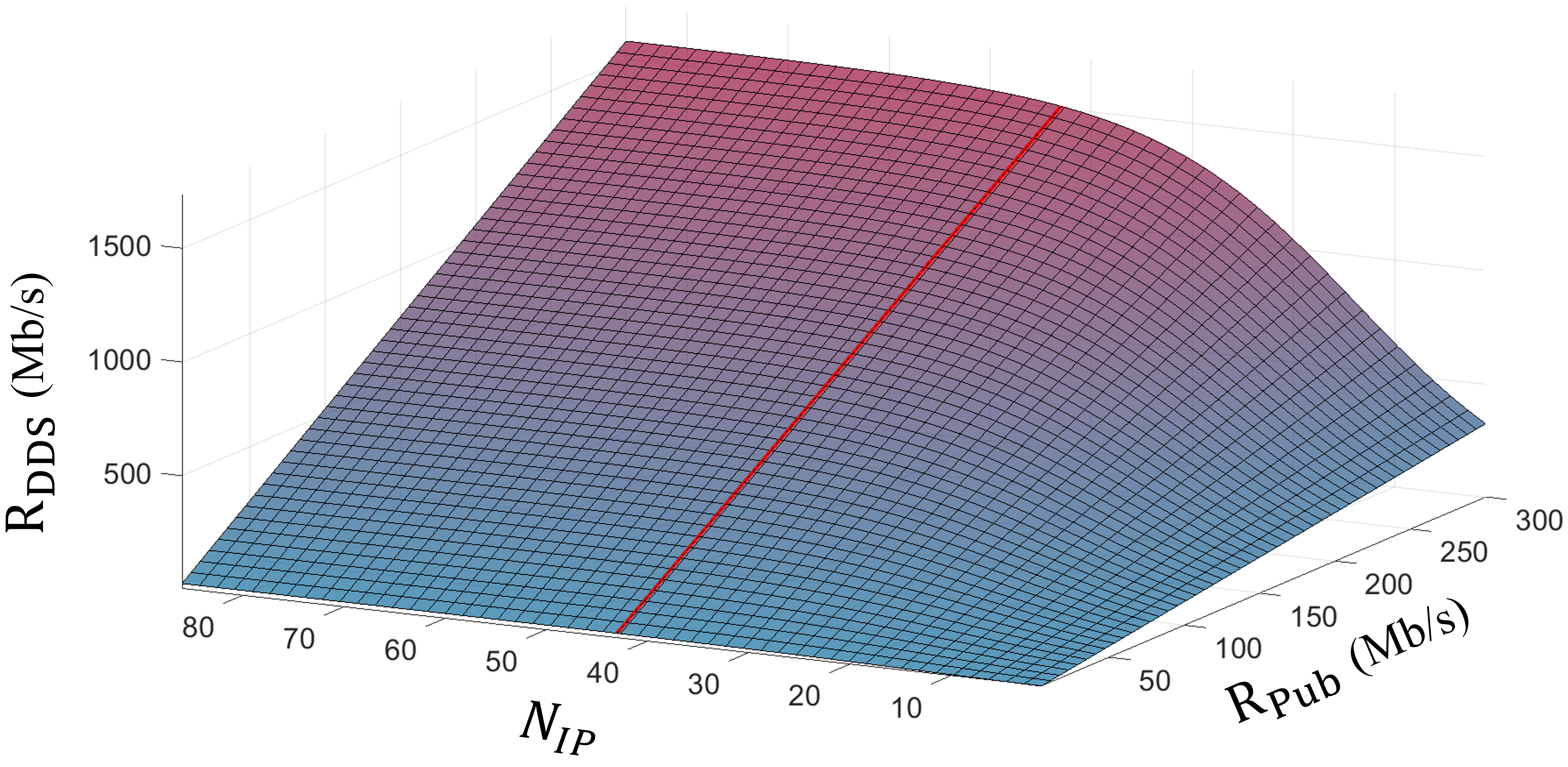}
    \caption{Impact of payload size and $N_{\mathrm{IP}}$ on $R_{\mathrm{DDS}}$ at PER 10\%.}
    \label{fig:Traffic}
\end{figure}
Given that retransmission rounds occur $n$ times per second, the steady-state DDS transmission rate, denoted by $R_{\mathrm{DDS}}$, is obtained by multiplying the per-round transmission $X$ by the retransmission rate $n$. 
\begin{equation}
\label{eq_mean}
R_{\mathrm{DDS}} = n \cdot X = \frac{ R_{\text{pub}}}{1 - p^2 \left(1 - p^{N_{\mathrm{IP}}} \right)}.
\end{equation}
This result demonstrates that the $R_{\mathrm{DDS}}$ is determined solely by $R_{\text{pub}}$, $p$, and $N_{\mathrm{IP}}$, and is independent of the retransmission rate $n$.

As shown in Eq.~\ref{eq_mean}, IP fragmentation quantified by $N_{\mathrm{IP}}$ is the dominant factor that amplifies transmission load. 
By default, DDS sets the maximum RTPS message size to 64~KB, while the typical IP MTU is 1500~B, resulting in $N_{\mathrm{IP}} = \lceil 64\,\text{KB} / 1500 \rceil = 44$.
Fig.~\ref{fig:Traffic} illustrates how $R_{\mathrm{DDS}}$ varies with respect to $R_{\text{pub}}$ and $N_{\mathrm{IP}}$, highlighting the severe overhead introduced by fragmentation in lossy networks.
As the payload size increases, $R_{\mathrm{DDS}}$ grows linearly. 
In contrast, reducing the maximum RTPS message size (i.e., decreasing $N_{\mathrm{IP}}$) causes an exponential decrease in $R_{\mathrm{DDS}}$. 
Therefore, adjusting the RTPS message size is significantly more effective in reducing traffic than lowering the application payload size. 

The most effective and straightforward way to mitigate this problem is to prevent IP fragmentation by enforcing a one-to-one mapping between an RTPS message and a single IP packet, i.e., $N_{\mathrm{IP}} = 1$. 
Most DDS vendors provide XML profile interfaces that allow users to configure the maximum RTPS message size. 
Given that the RTPS header size is 28~B and the standard MTU is 1500~B, we recommend setting the maximum RTPS message size to 1472~B to ensure fragmentation-free transmission at the network layer.
 
\begin{figure}[t!]
    \centering
    \includegraphics[width=\linewidth]{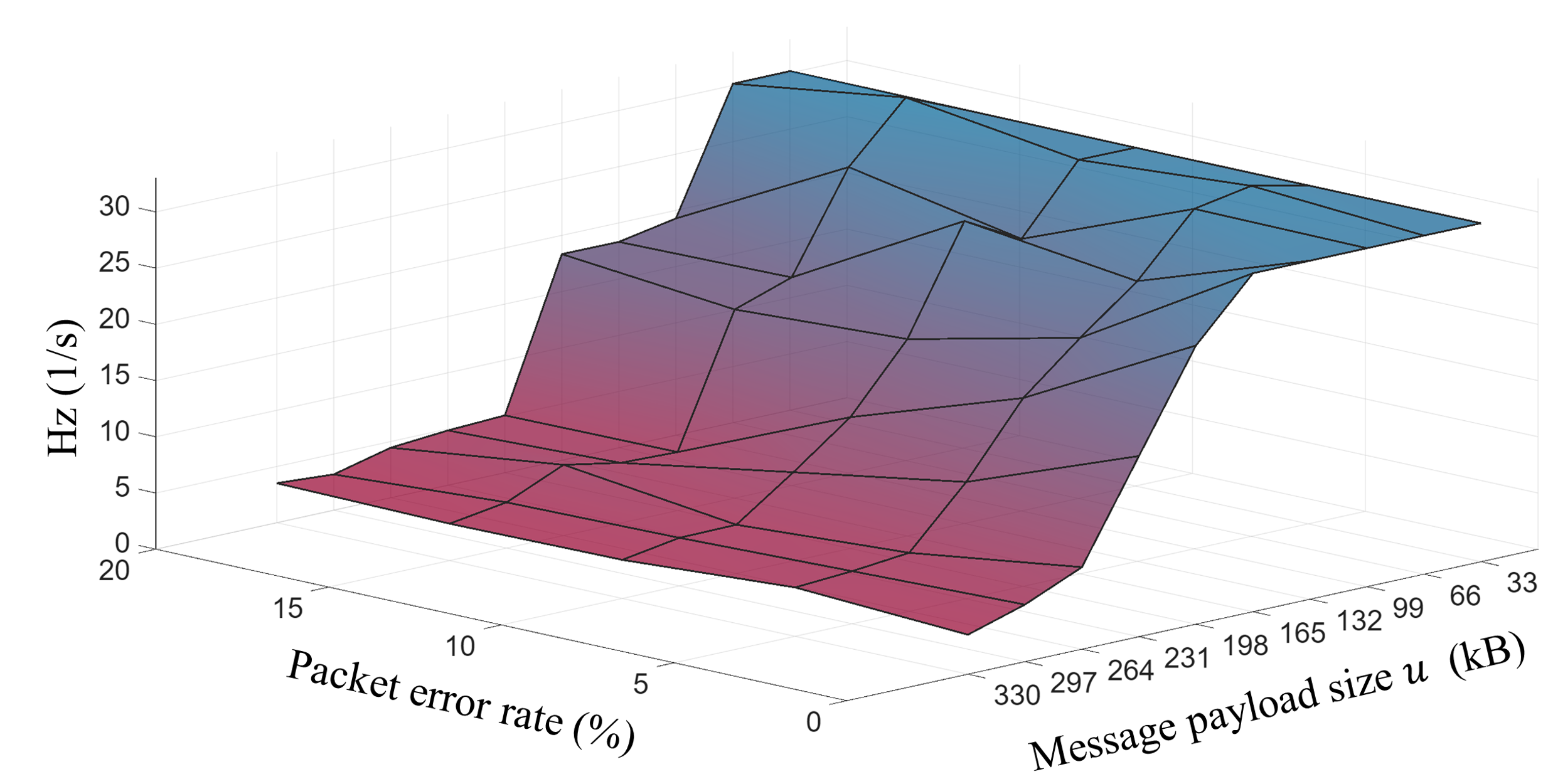}
    \caption{Effect of IP fragmentation prevention.}
    \label{fig:fragment}
\end{figure}

Fig.~\ref{fig:fragment} shows the subscriber-side reception rate (Hz) when publishing UInt8Array payloads at 30~Hz over IEEE~802.11ac.  
The results demonstrate the strong effectiveness of fragmentation-free transmission in lossy network conditions. 
Under the default configuration, message reception completely fails even with a modest PER, as demonstrated in Section~\ref{sec5-3}.
In contrast, when fragmentation prevention is enabled, reliable reception at 30~Hz is sustained up to PER 20\% for small payloads. 
Although message delivery never fully fails, the reception rate gradually decreases as PER and payload size increase, prompting the retransmission timing optimization discussed in the next section.

\subsection{Retransmission Timing}
\label{sec4-4}
Although the average transmission rate $R_{\mathrm{DDS}}$ is unaffected by the retransmission rate $n$ (Eq.~\ref{eq_mean}), the asynchronous interplay between publishing and retransmission makes the traffic inherently bursty, which can overload buffers and trigger link congestion.
To quantify the effect of burst traffic, we define the burst size $B_{\text{DDS}}$ as the number of bytes that leave the NIC's Tx ring back-to-back in a single retransmission round.
Since every RTPS message is immediately pushed to the Tx ring buffer, the worst-case burst arises when a new publication coincides with a retransmission.
It results in a burst that includes one fresh sample of size $u$ and the retransmission of backlog $X$ from the previous round. Hence,
\begin{equation}
\label{eq_burst}
    B_{\text{DDS}}=u+ p^2(1-q)\cdot X  \;=\;  u  \;+\; \frac{r\,u \, (p^2-p^2q)} {n\bigl(1-p^{2}+p^{2}q\bigr)}. 
\end{equation}

\begin{figure} [ht!]
    \centering
    \includegraphics[width=\linewidth]{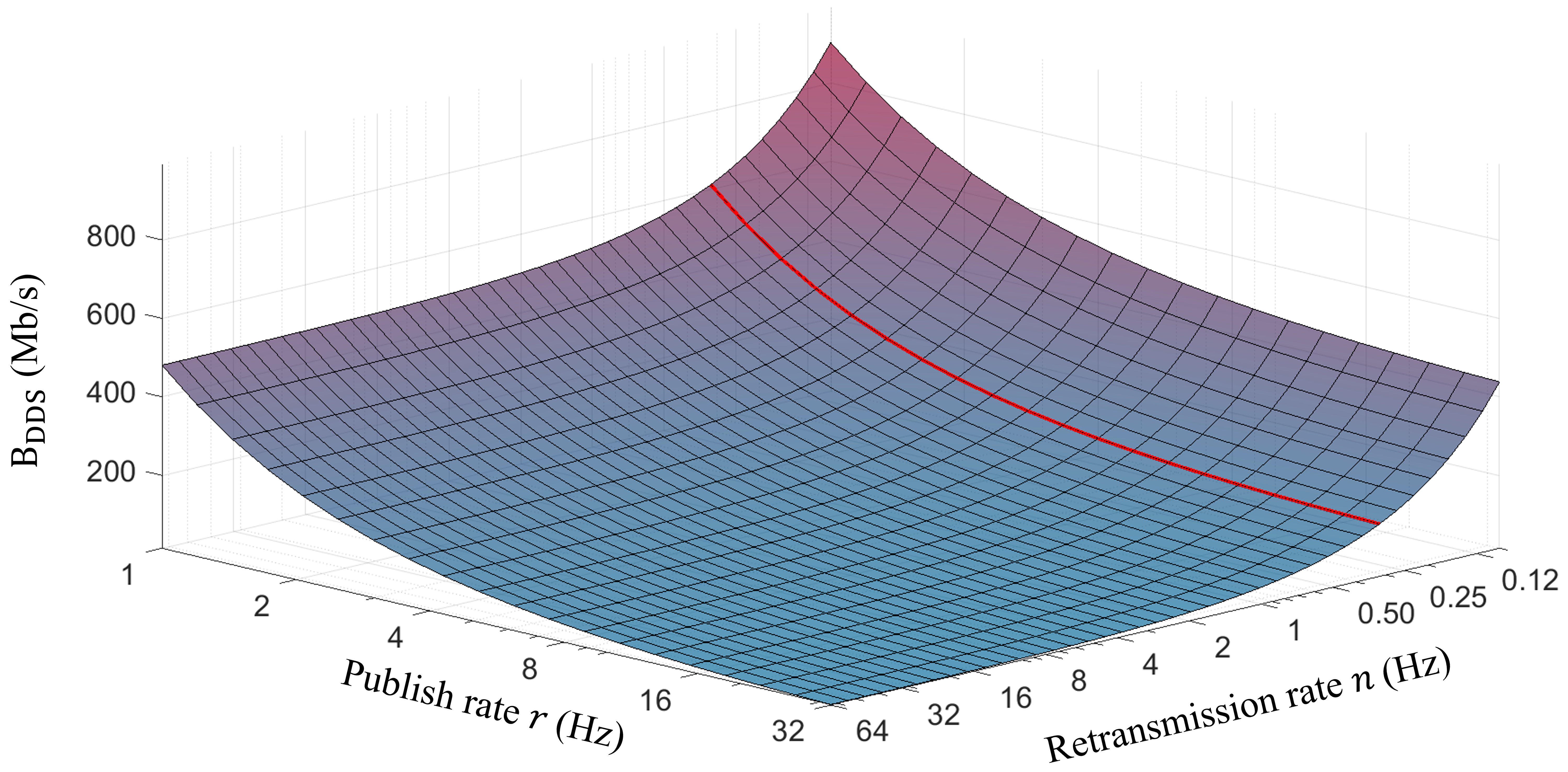}
    \caption{DDS burst traffic vs. publish and retransmission rate.}
    \label{fig:burst}
\end{figure}

Fig.~\ref{fig:burst} presents the relationship between DDS burst traffic $B_{\mathrm{DDS}}$, the publish rate $r$, and the retransmission rate $n$, under a fixed $R_{\text{pub}}$ of 480~Mb/s and a PER of 10\%.
The red line marks the default DDS retransmission rate of 0.33~Hz (i.e., a 3~s interval). 
As both $r$ and $n$ rates increase, $B_{\mathrm{DDS}}$ decreases significantly, indicating reduced burstiness. 
For example, publishing 65~KB messages at 30~Hz yields a sustained publish rate of only 16~Mb/s. However, under the default DDS configuration, a single retransmission event can trigger an instantaneous burst exceeding 160~Mb/s.
Such bursts can lead to buffer overflows and link congestion, resulting in higher latency and additional packet losses. 

These findings suggest that the default DDS retransmission rate is inappropriate for high-frequency data streams in real-time robotic applications. 
When the publish rate cannot be adjusted, shortening the retransmission rate is an effective means to suppress traffic bursts.
Increasing the retransmission rate $n$ can naturally reduce both average latency and jitter. Park \emph{et al.}~\cite{park2025analytical} demonstrated through probabilistic analysis that reducing the retransmission rate significantly improves latency characteristics in ROS~2 DDS, achieving exponential reductions in the mean delay and the jitter. 
However, excessively high values of $n$ may introduce overhead due to increased processing from the Heartbeat mechanism and a surge in RTPS control packets.
Lee \emph{et al.}~\cite{Submitted2025} identified a local optimum when setting the retransmission rate $n$ to $2r$, i.e., double the publish rate.
Based on these findings, we recommend setting $n = 2r$ to suppress retransmission-induced traffic bursts and minimize delay and jitter. 
This parameter can be easily configured via the QoS profile in DDS XML settings.

\begin{figure}[t!]
    \centering
    \includegraphics[width=\linewidth]{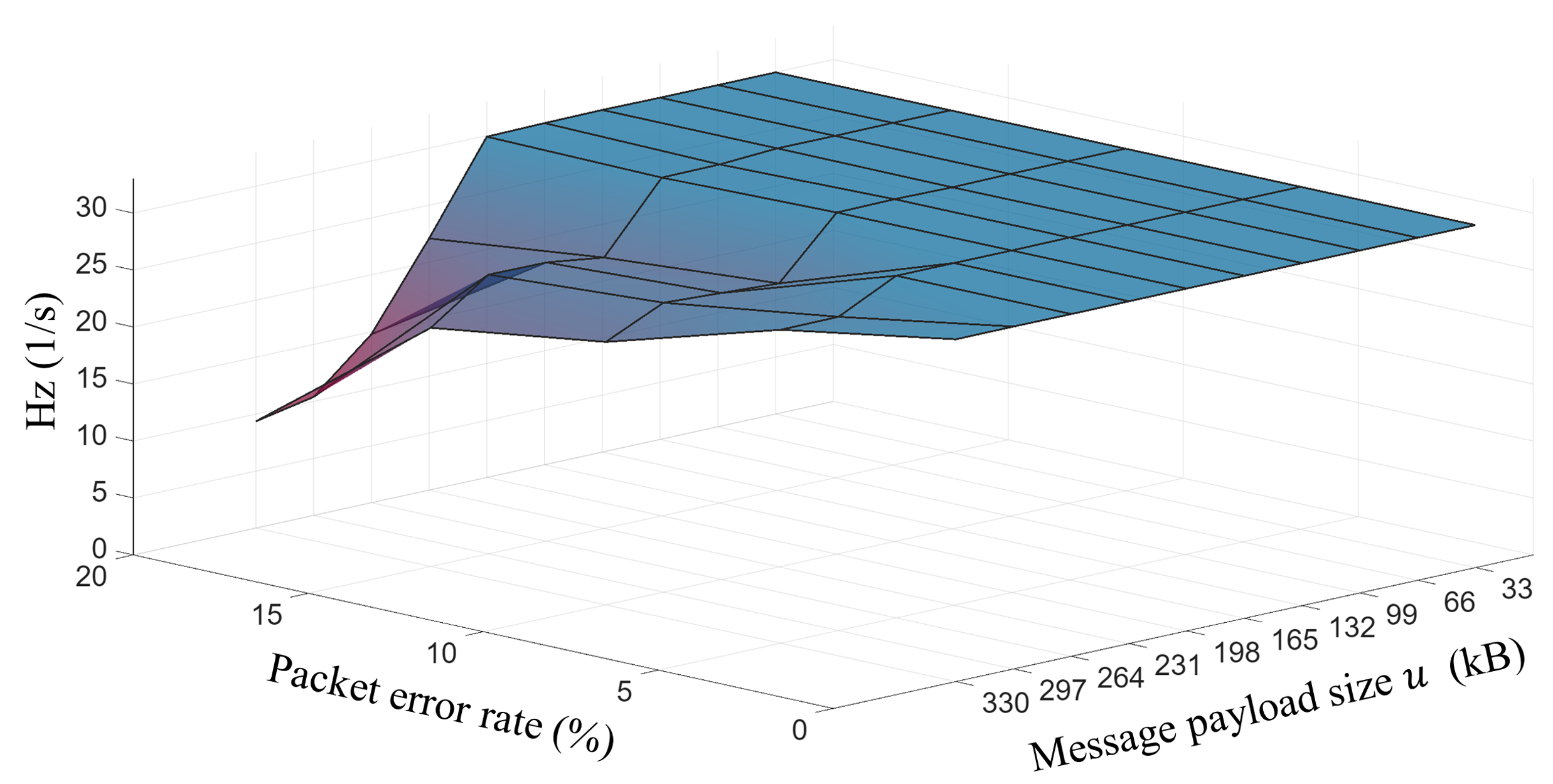}
    \caption{Effect of retransmission rate optimization.}
    \label{fig:retrans}
\end{figure}

Fig.~\ref{fig:retrans} illustrates the impact of the retransmission rate optimization, while all other conditions remain identical to Fig.~\ref{fig:fragment}.
The system maintains a stable reception rate of 30~Hz even with PER 20\% on highly lossy links, for payload sizes up to 231~KB. 
For larger payloads up to 330~KB, the reception rate remains at 30~Hz when no interference is present, but it gradually degrades as PER increases. 
This degradation results from the retransmission-induced traffic exceeding $T_{\text{OS}\rightarrow\text{Link}}$, which is inevitable.

\subsection{Buffer Burst}
\label{sec4-5}
Another critical challenge in wireless communication is the occurrence of link outages. 
Unlike wired networks, wireless links are not physically persistent and can be temporarily disconnected due to environmental obstructions, line-of-sight (LoS) loss, handovers, or movement outside the coverage area. 
When a link outage occurs, neither the ROS application nor the underlying DDS middleware is aware of the disruption. 
As a result, the ROS application continues publishing data, and DDS continues to store samples in the HistoryCache. 
This accumulation persists until the HistoryCache reaches its capacity. 
As described in Section~\ref{sec3-2}, DDS does not purge samples from the HistoryCache until acknowledgments are received from all expected subscribers. 
Consequently, during the link outage, the buffer becomes filled with unacknowledged samples.

This scenario exposes two fundamental limitations of DDS. 
First, unlike socket or ring buffers that are typically byte-based, the HistoryCache is size-limited by the number of samples. 
For large payloads, this results in significantly greater memory pressure under the same configuration. 
Second, the data rate from DDS to the OS layer, $T_{\text{DDS}\rightarrow\text{OS}}$ is typically much greater than $T_{\text{OS}\rightarrow\text{Link}}$.
And DDS performs no explicit flow control with respect to the available link capacity.

When the link is restored, DDS triggers a buffer burst. 
Specifically, upon receiving an AckNack from the DataReader, the DataWriter retransmits all stored samples in sequence from the HistoryCache. 
This sudden surge overwhelms the lower-layer buffers and saturates the wireless channel. 
In wireless networks, link saturation increases the packet error rate due to collisions, which in turn amplifies the retransmission load-creating a positive feedback loop that significantly increases transmission latency.

\begin{figure} [t!]
    \centering
    \includegraphics[width=0.7\linewidth]{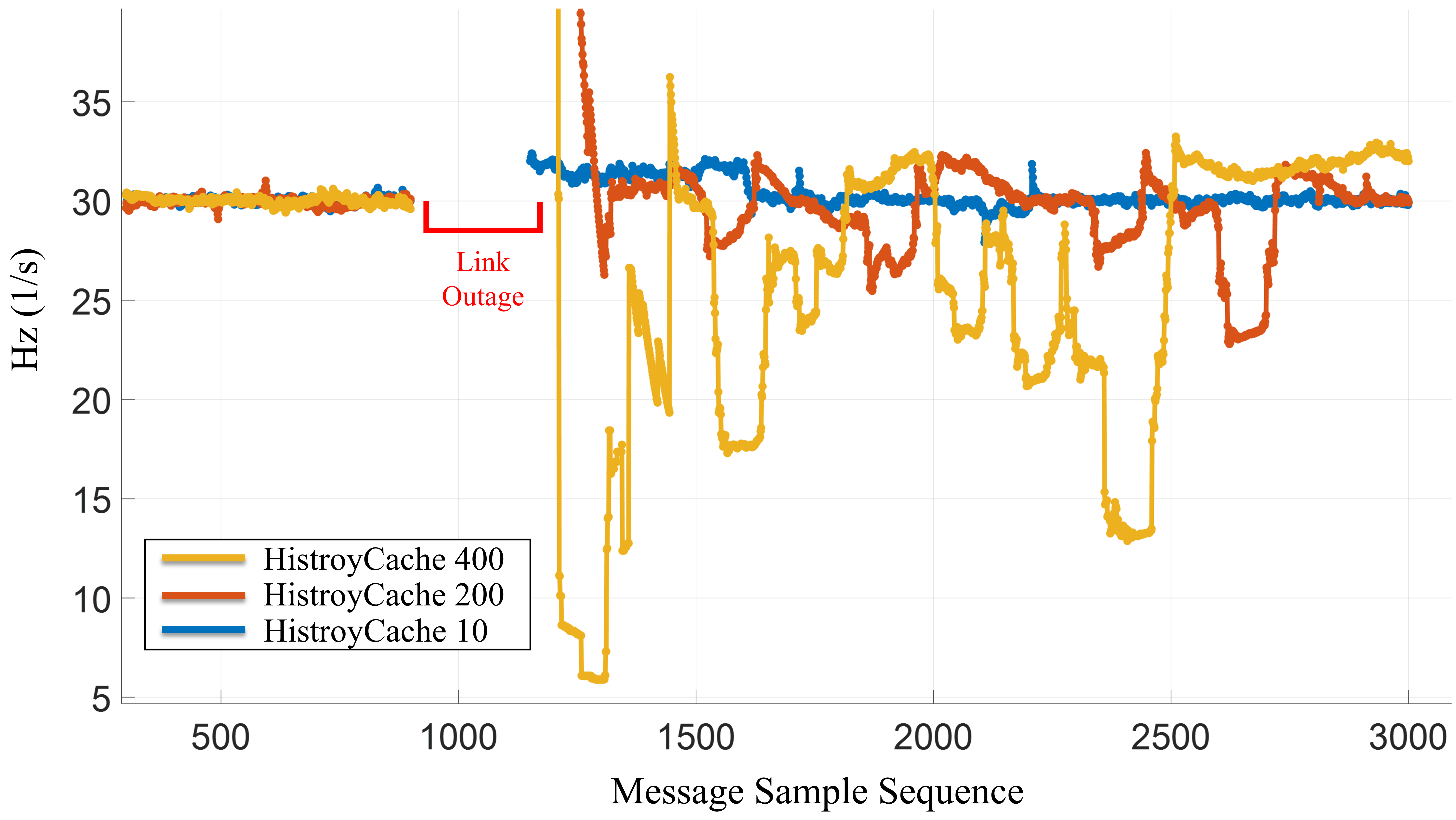}
    \caption{Buffer burst in link outage scenario.}
    \label{fig:buffer}
\end{figure}
Fig.~\ref{fig:buffer} presents the reception rate (Hz) in a scenario where a 10-second link outage occurs while transmitting 231~KB messages at 30~Hz. 
While the transmission rate remains stable at 30~Hz under normal conditions, it becomes unstable after the outage depending on the size of the HistoryCache. 
With a HistoryCache size of 10 (blue line), the system quickly returns to 30~Hz upon reconnection. 
However, with a size of 200 (orange line), the transmission rate fluctuates around 30~Hz for an extended period. 
At the ROS~2 default setting of 400 (yellow line), the transmission rate drops to as low as 5~Hz and remains degraded for a prolonged time, significantly reducing the effective throughput.
This results in an instantaneous buffer burst of up to $400 \times 231$~KB, forming a positive feedback loop that leads to collisions and further retransmissions.

The solution to prevent buffer bursts is to limit the size of HistoryCache through the DDS QoS parameter.
Given the payload size $u$ and $T_{\text{OS}\rightarrow\text{Link}}$, the optimized HistoryCache size $N_{\text{HC}}$ can be determined as
\begin{equation}
    N_{\text{HC}} = \left\lfloor \frac{T_{\text{OS}\rightarrow\text{Link}} \cdot \omega}{u} \right\rfloor,
\end{equation}
where $\omega$ denotes the link utilization allocated to DDS traffic. 
We recommend $\omega = 0.6$--$0.7$ when DDS occupies the link exclusively, and smaller values when multiple hosts share the channel.

\begin{figure}[t!]
    \centering
    \includegraphics[width=\linewidth]{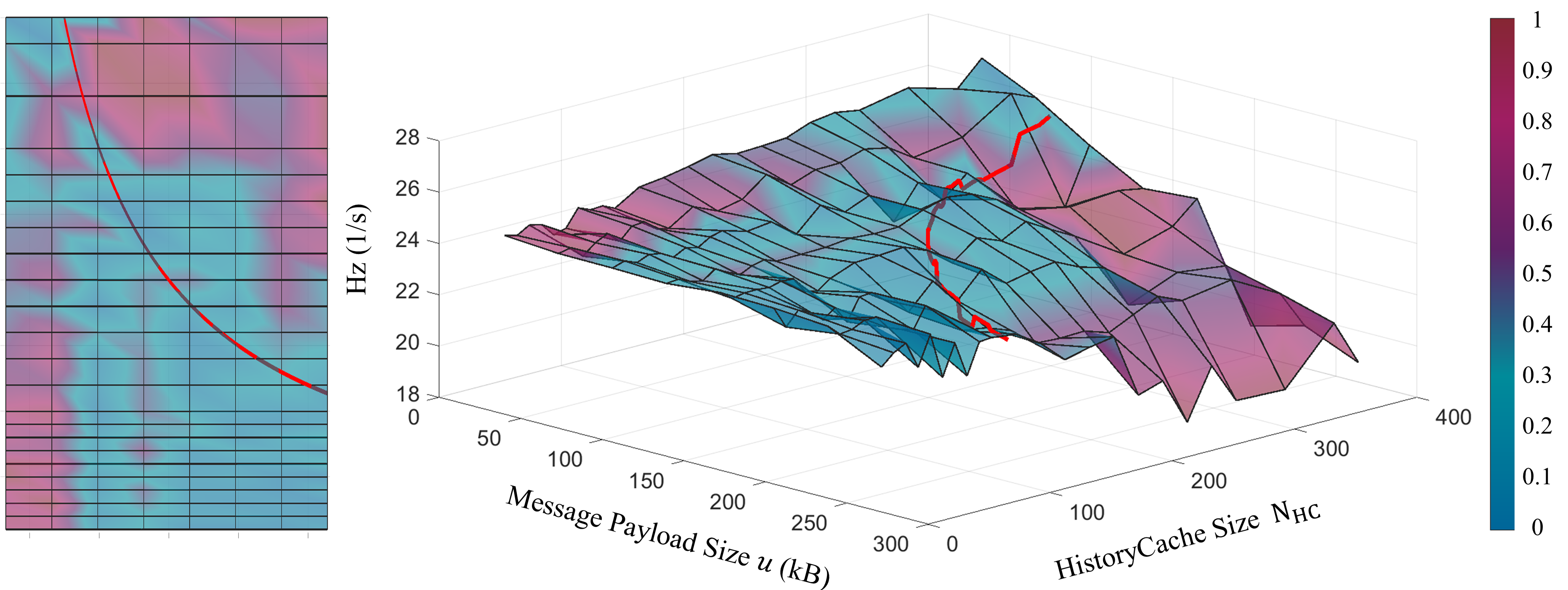}
    \caption{Effect of HistoryCache size optimization.}
    \label{fig:histcache}
\end{figure}
Fig.~\ref{fig:histcache} shows the reception rate (Hz) in a scenario where 30~Hz payloads of size $u$~KB are transmitted, followed by a 20-second link outage and recovery. 
The results illustrate the impact of HistoryCache size $N_{\text{HC}}$ on post-recovery performance, demonstrating the effectiveness of HistoryCache size optimization. 
The red vertical line indicates the optimal $N_{\text{HC}}$ based on the calculated threshold, assuming $T_{\text{OS}\rightarrow\text{Link}} \cdot \omega = 240$~Mb/s.

% To the left of this threshold, increasing $N_{\text{HC}}$ improves the reception rate by enabling faster recovery from backlog. 
Beyond the threhold, larger $N_{\text{HC}}$ values cause a sharp degradation in reception rate. 
This is because moderate buffer bursts, within link capacity, allow the retransmission backlog to be flushed efficiently, whereas excessive bursts exceed link capacity, leading to congestion and reduced reception.
These results highlight that, given only $u$ and $T_{\text{OS}\rightarrow\text{Link}}$, one can determine the optimal $N_{\text{HC}}$ that maximizes post-outage performance without saturating the link.

\subsection{DDS Optimization Framework}
\label{sec4-6}
We propose a DDS optimization framework that simultaneously addresses all three key failure causes identified in previous sections. 
The framework is entirely based on the standard ROS~2 XML-based QoS configuration interface, requiring no changes to the underlying protocol, additional implementation, or complex integration procedures. 
It is designed to be immediately applicable to any ROS~2 application with minimal setup, serving as a robust and lightweight solution for wireless communication optimization.

\begin{tcolorbox}[
  title=DDS Optimization for Wireless Large Payload Transfer,
  colback=white,              
  colframe=black,            
  boxrule=0.5pt,             
  sharp corners,             
  before upper=\setlength{\parindent}{0pt}
]
\textbf{Input:} Publish rate $r$, payload size $u$, link-layer throughput $T_{\text{OS}\rightarrow\text{Link}}$, link utilization $\omega$ \\
\textbf{Step 1:} Set RTPS maxMessageSize = 1472~B \\
\textbf{Step 2:} Set retransmission rate $n = 2r$ \\
\textbf{Step 3:} Set HistoryCache size as \\
\[
N_{\text{HC}} = \left\lfloor \frac{T_{\text{OS}\rightarrow\text{Link}} \cdot \omega}{u} \right\rfloor
\]
\textbf{Output:} Optimized ROS~2 XML QoS profile
\end{tcolorbox}

Detailed usage instructions and source code are available at our open-source repository: \href{https://github.com/csi-dgist/DDS-Optimizer-for-Wirelss-Large-Payload-Transfer}{\texttt{[Github Link]}}.
Users can apply wireless DDS optimization to any ROS~2 application by specifying only four input parameters, without modifying application logic or rebuilding the system.
In the next section, we demonstrate the effectiveness of the proposed DDS optimization through extensive empirical evaluation in various wireless link and traffic scenarios.

\section{Experimental Evaluation}
\label{sec5}
In this section, we evaluate the robustness and generalizability of the proposed DDS optimization across diverse wireless conditions and payload sizes.
We compare it against two baselines: (i) the default DDS configuration and (ii) Fast DDS's LARGE\_DATA mode.

\subsection{Experimental Setup}
\label{sec5-1}
All experiments are conducted using two laptops connected via an IEEE~802.11ac 5~GHz wireless router. 
The theoretical throughput of the link, denoted as $T_{\text{OS}\rightarrow\text{Link}}$, is 433~Mb/s. 
Each laptop is equipped with an Intel Cannon Point-LP CNVi wireless NIC and runs Ubuntu~22.04 with ROS~2 Humble. 
Fast DDS v2.6.9 is used as the default DDS implementation.
To emulate a typical ROS~2 user experience, we perform all tests without modifying any DDS or ROS source code, and leave all OS kernel and transport-layer settings at their defaults. 
One laptop operates as a publisher and the other as a subscriber.

\subsection{Experimental Scenarios}
\label{sec5-2}
We consider four distinct wireless conditions:

\begin{itemize}
  \item \textbf{Ideal link:} clean channel with no external interference. DDS traffic exclusively occupies the wireless link.
  \item \textbf{Mild loss:} PER 1\% is injected to simulate moderate background noise. This is achieved by dropping packets at the NIC level using Linux traffic control tools.
  \item \textbf{Severe loss:} PER 20\% is injected to emulate environments with high interference or physical signal degradation.
  \item \textbf{Link outage:} 20-second disconnection is enforced mid-transmission, after which the link is restored.
\end{itemize}

For each scenario, we transmit payloads of 32, 64, 128, 256, and 512~KB at 30~Hz using ROS~2's rclpy interface. 
We transmit standard ROS~2 message types, images point clouds, string, and UInt8Array. 
All transmissions are configured with RELIABLE reliability and KEEP\_ALL history to ensure strict reliability.

For each configuration, 1,000 messages are transmitted in five independent trials per message type.
After completing all trials across all message types, we compute the overall average for each metric.
We measure the reception rate (Hz), defined as the number of received messages divided by the time span between the first and the last message, along with the average latency and the jitter (i.e., the standard deviation of latency).

\subsection{Experimental Results}
\label{sec5-3}
Table~\ref{table:performance} summarizes the experimental results. 
Subtables (a), (b), (c), and (d) correspond to the ideal link, mild loss, severe loss, and link outage scenarios, respectively. 
A dash (---) in reception rate indicates that communication was not completed and the process overran. 
In cases where the average latency exceeded 3,300~ms (100 times the transmission interval), both the average latency and jitter are represented as ``---''.
\begin{table*}[htbp]
\centering
\caption{Performance comparison across DDS modes and payload sizes}
\label{table:performance}
\begin{subtable}[t]{\textwidth}
\centering
\caption{Ideal Link Condition}
\label{ideal}
\small
\begin{tabular}{l*{15}{>{\centering\arraybackslash}p{1.7em}}}
\toprule
\textbf{Mode} & \multicolumn{5}{c}{Default DDS Profile} & \multicolumn{5}{c}{LARGE\_DATA Mode} & \multicolumn{5}{c}{Optimized DDS Profile} \\
\cmidrule(lr){2-6} \cmidrule(lr){7-11} \cmidrule(lr){12-16}
\textbf{Payload (Kb)} & 32 & 64 & 128 & 256 & 512 & 32 & 64 & 128 & 256 & 512 & 32 & 64 & 128 & 256 & 512 \\
\midrule
Reception rate (Hz) & 30.09 & 30.08 & 30.08 & 29.36 & 1.16 & 29.15 & 18.40 & 17.29 & 11.98 & 8.54 & 30.08 & 30.08 & 30.08 & 30.07 & 28.89 \\
Avg. Latency (ms)   & 5.47 & 9.33 & 10.97 & 33.94 & --- & 4.86 & 110.82 & 78.65 & 93.22 & 130.99 & 4.59 & 6.30 & 7.69 & 12.78 & 56.25 \\
Jitter              & 2.51 & 4.36 & 10.02 & 57.59 & --- & 1.47 & 177.51 & 173.83 & 132.34 & 43.45 & 1.59 & 2.19 & 3.42 & 2.39 & 109.32 \\
\bottomrule
\end{tabular}
\end{subtable}

\vspace{1em}

\begin{subtable}[t]{\textwidth}
\centering
\caption{Mild Loss Condition}
\label{mild}
\small
\begin{tabular}{l*{15}{>{\centering\arraybackslash}p{1.7em}}}
\toprule
\textbf{Mode} & \multicolumn{5}{c}{Default DDS Profile} & \multicolumn{5}{c}{LARGE\_DATA Mode} & \multicolumn{5}{c}{Optimized DDS Profile} \\
\cmidrule(lr){2-6} \cmidrule(lr){7-11} \cmidrule(lr){12-16}
\textbf{Payload (Kb)} & 32 & 64 & 128 & 256 & 512 & 32 & 64 & 128 & 256 & 512 & 32 & 64 & 128 & 256 & 512 \\
\midrule
Reception rate (Hz) & 1.33 & 0.94 & 0.15 & --- & --- & 19.52 & 14.15 & 7.28 & 5.84 & 5.86 & 30.08 & 30.05 & 30.02 & 29.99 & 21.86 \\
Avg. Latency (ms)   & --- & --- & --- & --- & --- & 129.92 & 167.03 & 253.03 & 229.86 & 201.12 & 7.77 & 11.49 & 14.71 & 29.50 & 477.56 \\
Jitter              & --- & --- & --- & --- & --- & 220.76 & 270.11 & 289.76 & 314.59 & 323.89 & 7.00 & 8.35 & 7.03 & 18.18 & 358.16 \\
\bottomrule
\end{tabular}
\end{subtable}

\vspace{1em}

\begin{subtable}[t]{\textwidth}
\centering
\caption{Severe Loss Condition}
\label{severe}
\small
\begin{tabular}{l*{15}{>{\centering\arraybackslash}p{1.7em}}}
\toprule
\textbf{Mode} & \multicolumn{5}{c}{Default DDS Profile} & \multicolumn{5}{c}{LARGE\_DATA Mode} & \multicolumn{5}{c}{Optimized DDS Profile} \\
\cmidrule(lr){2-6} \cmidrule(lr){7-11} \cmidrule(lr){12-16}
\textbf{Payload (Kb)} & 32 & 64 & 128 & 256 & 512 & 32 & 64 & 128 & 256 & 512 & 32 & 64 & 128 & 256 & 512 \\
\midrule
Reception rate (Hz) & --- & --- & --- & --- & --- & 0.65 & 0.25 & 0.10 & 0.05 & 0.02 & 30.10 & 30.07 & 30.03 & 30.04 & 12.91 \\
Avg. Latency (ms)   & --- & --- & --- & --- & --- & --- & --- & --- & --- & --- & 2.27 & 14.79 & 22.24 & 44.25 & 1300.60 \\
Jitter              & --- & --- & --- & --- & --- & --- & --- & --- & --- & --- & 8.33 & 10.13 & 10.13 & 19.47 & 701.73 \\
\bottomrule
\end{tabular}
\end{subtable}

\vspace{1em}

\begin{subtable}[t]{\textwidth}
\centering
\caption{Link Outage Condition}
\label{link}
\small
\begin{tabular}{l*{15}{>{\centering\arraybackslash}p{1.7em}}}
\toprule
\textbf{Mode} & \multicolumn{5}{c}{Default DDS Profile} & \multicolumn{5}{c}{LARGE\_DATA Mode} & \multicolumn{5}{c}{Optimized DDS Profile} \\
\cmidrule(lr){2-6} \cmidrule(lr){7-11} \cmidrule(lr){12-16}
\textbf{Payload (Kb)} & 32 & 64 & 128 & 256 & 512 & 32 & 64 & 128 & 256 & 512 & 32 & 64 & 128 & 256 & 512 \\
\midrule
Reception rate (Hz) & --- & --- & --- & --- & --- & 16.23 & 14.92 & 14.85 & 13.01 & 6.42 & 26.81 & 24.22 & 25.01 & 24.20 & 10.57 \\
Avg. Latency (ms)   & --- & --- & --- & --- & --- & 25.25 & 38.73 & 41.48 & 69.33 & 146.24 & 14.61 & 20.12 & 26.58 & 33.82 & 306.38 \\
Jitter              & --- & --- & --- & --- & --- & 71.96 & 104.36 & 71.26 & 72.28 & 84.03 & 53.41 & 51.86 & 21.85 & 20.69 & 266.84 \\
\bottomrule
\end{tabular}
\end{subtable}
\end{table*}

Table~\ref{ideal} shows the transmission performance under ideal link conditions.
Interestingly, the Default DDS Profile outperformed the LARGE\_DATA Mode across most payload sizes.
Except for the 512~KB case, it achieved higher reception rates and lower latency, indicating that UDP-based transmission can outperform TCP when bandwidth is sufficient and the link is stable.
While the Default profile degraded sharply with increasing payload size, the LARGE\_DATA Mode, though competitive at 32~KB, showed consistent declines across all metrics as payloads grew-likely due to fragmentation, retransmission, and header overhead.

In contrast, the Optimized DDS Profile sustained a high reception rate of 28.9~Hz and a low latency of 56~ms even at 512~KB.
This shows that the proposed optimization efficiently utilizes bandwidth by reducing unnecessary retransmissions, achieving the lowest latency and jitter among all profiles and proving its advantage in real-time wireless communication.

% Table~\ref{ideal} presents the transmission performance under an ideal link condition.
% Interestingly, the Default DDS Profile consistently outperformed the LARGE\_DATA Mode across most payload sizes.
% Except for the 512~KB case, the default configuration achieved higher reception rates and lower latencies, suggesting that UDP-based transmission in DDS can deliver superior performance over TCP when the network bandwidth is abundant and \textcolor{red}{conditions are stable}.
% While the default profile exhibited a sharp performance degradation as payload size increased, the LARGE\_DATA Mode, despite showing competitive performance at 32~KB, suffered from consistent degradation across all metrics as payloads grew larger.
% This implies that the TCP-based mode encounters bottlenecks when handling large payloads, likely due to fragmentation, retransmission, and header overhead.

% In contrast, the Optimized DDS Profile maintained a high reception rate of 28.9~Hz and a low average latency of 56~ms, even for the largest payload (512~KB).
% This demonstrates that the proposed optimization effectively utilizes link bandwidth, even in ideal conditions, by minimizing unnecessary retransmissions.
% As a result, it achieved the lowest average latency and jitter among all profiles, highlighting its superiority in real-time data delivery compared to both the Default and LARGE\_DATA configurations.

Table~\ref{mild} presents the transmission performance under a mild loss condition (PER = 1\%).
This scenario exposes the fragility of the Default DDS profile and the need for the LARGE\_DATA Mode.
Even under minor packet loss, the Default DDS profile becomes nearly non-functional-reception rates approach zero, and large payloads trigger premature process termination due to resource exhaustion.
This highlights the vulnerability of fragmentation in lossy networks.

The LARGE\_DATA Mode reliably delivers data across all cases, demonstrating the effectiveness of TCP's reliability mechanisms under moderate loss.
However, as payload size increases, reception rates degrade, and both latency and jitter rise sharply due to retransmissions and congestion control overhead.

The Optimized DDS Profile shows strong resilience to mild loss.
Up to 256~KB, it maintains high reception rates with minimal latency increase, confirming that optimized retransmission timing effectively reduces delay while preserving throughput.

At 512~KB, however, performance drops sharply.
This indicates that when the offered rate exceeds the effective bandwidth, even modest channel noise can severely impact transmission.
It underscores the importance of aligning the transmission rate $R_{\text{Pub}}$ with available bandwidth-a key design constraint often overlooked under ideal conditions.

Table~\ref{severe} shows the performance of each DDS mode under a severe loss condition (PER = 20\%).
While such high loss is uncommon, robustness under worst-case scenarios is essential for robotic systems.
Here, both the Default and LARGE\_DATA modes fail entirely, mainly due to the exponential retransmission overhead caused by fragmentation.

In contrast, the Optimized DDS Profile maintains stable transmission up to 256~KB, even under heavy loss.
This confirms that as long as $R_{\text{Pub}}$ remains within the effective bandwidth, efficient retransmission control enables sustained performance despite severe loss.

% Table~\ref{severe} presents the performance of each DDS mode under a severe loss condition of PER = 20\%.
% Although such extreme packet loss is rare in typical environments, ensuring communication robustness under worst-case scenarios is critical for robotic systems.
% The results are straightforward: neither the Default DDS mode nor the LARGE\_DATA mode functions under this condition.
% The primary reason lies in the exponential increase in retransmission traffic caused by fragmentation.

% In contrast, the Optimized DDS Profile maintains stable transmission up to 256~KB payloads, even under heavy loss.
% This confirms that as long as the transmission rate $R_{\text{Pub}}$ stays within the effective bandwidth, the system can sustain performance through efficient retransmission control, despite high packet loss.

Table~\ref{link} presents the performance of each DDS mode under a link outage condition.
The Default DDS Profile shows critical vulnerability: after a link is lost and later recovered, the system fails to recover due to buffer overruns and incomplete fragment retransmission. This stems from inefficient retransmission and excessive fragmentation, which overwhelm the link.

In contrast, the TCP-based LARGE\_DATA Mode employs flow and congestion control, enabling recovery after link restoration. However, its conservative nature slows transmission, degrading throughput and increasing latency and jitter.
The Optimized DDS Profile balances UDP throughput with controlled bandwidth via HistoryCache optimization, maintaining fast and low-latency transmission across all payloads except 512~KB-even under link disruption.

These findings reinforce that the proposed optimization strategy outperforms existing profiles across a wide range of network conditions. Importantly, it demonstrates that by leveraging network-aware design-particularly optimizing retransmission rate and history caching-even a lightweight UDP-based DDS stack can provide robust and efficient communication without relying on the complexity of TCP.

\section{Conclusion}
\label{sec6}
In this paper, we presented the first in-depth network-layer analysis of the ROS~2 DDS communication stack under wireless conditions with large payloads.
Through systematic modeling, we identified three key factors that hinder reliable data transmission: excessive IP fragmentation, inefficient retransmission timing, and DDS buffer bursts.
Based on these insights, we proposed a lightweight and fully compatible DDS optimization framework that enhances transmission robustness without requiring any changes to the protocol or application logic.
Our solution leverages simple XML-based QoS configurations and can be seamlessly integrated into existing ROS~2 deployments.

Extensive experimental results across diverse wireless scenarios demonstrate that the proposed optimized profile significantly improves performance in terms of reception rate, average latency, and jitter-particularly under adverse conditions such as packet loss and link outages.
Users expect ROS~2 to just work, even in the presence of complex networking challenges.
Our results show that with network-aware DDS tuning, robust and real-time wireless communication is achievable, paving the way for scalable, edge-driven robotic systems.

This study focuses on optimizing periodic communication in ROS~2 applications, specifically targeting regularly timed message flows. However, many real-world robotic systems rely on event-driven or aperiodic transmissions. Extending the proposed framework to handle such asynchronous communication patterns is an important direction for future work. Moreover, our current analysis assumes a one-to-one Pub-Sub communication model. To fully address the challenges of wireless robotics, future research should explore network-aware DDS optimization strategies in multi-node environments, where multiple data streams interact over shared physical links.

\newpage
\bibliographystyle{ieeetr}
\bibliography{biblist}
\end{document}